\documentstyle [12pt]{article}
\def\int {\intop \limits}

\def\fnote#1{\footnote}
\begin{document}
\newcommand{\dst}[1]{\displaystyle{#1}}
\newcommand{\barl}{\begin{array}{rl}}
\newcommand{\ball}{\begin{array}{ll}}
\newcommand{\ear}{\end{array}}
\newcommand{\barc}{\begin{array}{c}}
\newcommand{\sne}[1]{\displaystyle{\sum _{#1} }}
\newcommand{\sn}[1]{\displaystyle{\sum ^{\infty }_{#1} }}
\newcommand{\ini}[1]{\displaystyle{\int ^{\infty }_{#1}}}
\newcommand{\myi}[2]{\displaystyle{\int ^{#1}_{#2}}}
\newcommand{\inn}{\displaystyle{\int }}
\newcommand{\be}{\begin{equation}}
\newcommand{\ee}{\end{equation}}
\newcommand{\aq}[1]{\label{#1}}

\vspace*{4.0cm}
\centerline{\Large {\bf Transition radiation as a source of}}
\vskip .25cm
\centerline{\Large {\bf quasi-monochromatic X-rays}}
\vskip .5cm
\centerline{\large{\bf V. N. Baier and V. M. Katkov}}
\centerline{Budker Institute of Nuclear Physics,
 630090 Novosibirsk, Russia}
\vskip 2.0cm
\begin{abstract}
Transition radiation (TR) from ultrarelativistic particles is
considered. It is shown that performing collimation of the TR from
the periodic N-foil stack (parameters of which are selected in a
appropriate manner) one obtains the spectrum of the TR which has a form of
a peak position of which $\omega_1$ depends on the plasma
frequency and the thickness of the radiator foils.  The height and
width of the peak depend on the collimation angle $\vartheta_c$.
The height of the peak for given $\vartheta_c$ is proportional to N.
Selecting parameters one can have the source
of X-rays of desired frequency with rather good monochromaticity.
\end{abstract}

\newpage

\section{Introduction}

The transition radiation arises at
uniform and rectilinear motion of a charged
particle when it intersects a boundary of two different media (in general
case, when it moving in a nonuniform medium or near such medium).
This phenomenon \cite{1} was actively investigated
during a few last decades (see, e.g. reviews \cite{2}, \cite{3}) and
widely used in transition radiation detectors (see, e.g. recent review
\cite{4} and references therein).

We consider the standard TR radiator of $N$ foils of thickness $l_1$
separated by distances $l_2$ in a gas or the vacuum,
so the period length is $l=l_1+l_2$. The plasma frequency of the foil
and gap material are
$\omega_0$ and $\omega_{02}$, we neglect $\omega_{02}$.
The basic features of the TR in this radiator depend essentially on
the interrelation between values of
$\displaystyle{\omega_1=\frac{\omega_0^2 l_1}{4\pi}}$ and
$\displaystyle{\overline{\omega_p}=\omega_0 \gamma \sqrt{\frac{l_1}{l}}}$,
$\gamma$ is the Lorentz factor. In the TR detectors the inequality
$\omega_1 \gg \overline{\omega_p}$ is fulfilled, in the usual
case $l_1 \ll l_2$ and radiated frequencies $\omega < \omega_0 \gamma$.
The total energy radiated is proportional to $\gamma$ and the TR detector
are just used to measure this quantity.

In the opposite case $\omega_1 \ll \overline{\omega_p}$ and $l_1 \sim l_2$
characteristics of the TR are quite different. The total radiated energy
is independent of $\gamma$.
Performing the collimation of the radiation within angle
$\vartheta_c$ with respect velocity of the initial particle one can obtain
the radiation concentrated in a rather narrow spectral band near $\omega_1$.
The width of this band depend on $\vartheta_c$.
There are limitations on the number of foils $N$ due to
absorption of the radiated X-rays and multiple scattering of the projectile.
Nevertheless one can pick out parameters which permit obtain number of
radiated per one crossing photons $N_{\gamma} \sim 0.01 \div 1$
(per one projectile). This case is discussed in detail in Sec.2.

In Sec.3 the present situation with use of the TR
as a X-ray source is discussed.
Some specific features of the proposed radiator are analyzed
including selection of the parameters. Set of examples of
X-ray sources with various $\omega_1$ utilizing
distinct material of the foils and operating at different energies
is collected in Table.

\section{Transition radiation from the periodic N-foil stack}

The spectral-angular distribution of emitted from ultrarelativistic
electrons energy in the radiator
consisting of many ($N$) thin foils of the thickness $l_1$ separated by equal
distances $l_2$ in a gas was discussed in many papers,
see e.g. \cite{3}-\cite{7}
\begin{equation}
\frac{d^2\varepsilon}{d\omega dy}=\frac{4e^2y}{\pi}
\left(\frac{\kappa_0^2}{(1+y)(1+\kappa_0^2+y)} \right)^2
\sin^2\frac{\varphi_1}{2}~\frac{\sin^2 (N\varphi/2)}{\sin^2 (\varphi/2)},
\label{1}
\end{equation}
where $\omega$ is the frequency of radiation, $y=\vartheta^2 \gamma^2$,
$\gamma=\epsilon/m$ is the Lorentz factor, $\epsilon (m)$ is the energy
(the mass) of the incident electron, $\vartheta$ is the azimuthal angle
of emission with respect velocity of the incident electron (we assume
normal incidence), $\kappa_0=\omega_p/\omega$,
here $\omega_p=\omega_0\gamma$, $\omega_0$ is the plasma frequency
\begin{equation}
\omega_0^2=\frac{4\pi e^2n_e}{m},\quad \varphi_1=\frac{\omega l_1}{2\gamma^2}
\left(1+\kappa_0^2+y \right),\quad
\varphi_2=\frac{\omega l_2}{2\gamma^2}
\left(1+y \right),\quad \varphi=\varphi_1+\varphi_2,
\label{2}
\end{equation}
where $n_e$ is the density of electrons in the medium of a foil.
The radiated energy is the coherent sum of the TR
amplitudes for each interface
and in absence of absorption Eq.(\ref{1}) has the pronounced
interference pattern.

Although the formula (\ref{1}) is derived in classical electrodynamics,
one can introduce the probability of the TR
\begin{equation}
\frac{d^2w}{d\omega dy}=\frac{1}{\omega} \frac{d^2\varepsilon}{d\omega dy}
\label{3}
\end{equation}
In this paper the system $\hbar=c=1$ is used, $e^2=\alpha=1/137$.
Recently authors developed the quantum theory of the TR and of the transition
pair creation \cite{6a}.

At $N \gg 1$ the main contribution into the integral over $y$ in (\ref{1}),
which defines the spectral distribution of the TR, gives the interval of
$\varphi$ for which
\begin{equation}
\sin^2 \varphi/2 \ll 1,\quad \varphi=2\pi n+
\Delta \varphi,\quad \Delta \varphi \sim \frac{1}{N},\quad
\Delta y \sim \frac{1}{N}\frac{2\gamma^2}{\omega l}=\frac{1}{N}\frac{l_c}{l},
\label{4}
\end{equation}
where $l_c=2\gamma^2/\omega$ is the formation length of radiation in
the vacuum, $l=l_1+l_2$. The condition
\begin{equation}
\varphi=\varphi_1+\varphi_2=\frac{\omega l}{2\gamma^2}
\left(1+y \right)+\frac{\omega l_1}{2\gamma^2}\kappa_0^2=2\pi n
\label{5}
\end{equation}
defines the radiated photon energy $\omega$ as a function of
the emission angle $\vartheta$ for fixed $n$ (or for the $n$-th radiation
harmonic $\omega_n$). Respectively, the integral over $y$ in (\ref{1})
can be presented as a sum of the harmonic. We present (\ref{5}) in a form
\begin{equation}
1+y=\frac{l_1}{l}\left(\frac{\omega_p}{\omega_n}\right)^2
\left(1-\frac{\omega_n}{\omega} \right) \frac{\omega_n}{\omega}=
\frac{\overline{\omega_p^2}}{\omega_n^2}
\left(1-\frac{\omega_n}{\omega} \right) \frac{\omega_n}{\omega},
\label{6}
\end{equation}
where
\[
\omega_n=\frac{\omega_1}{n},\quad \omega_1=\frac{\omega_0^2 l_1}{4\pi},\quad
\overline{\omega_p^2}=\gamma^2 \overline{\omega_0^2}=\gamma^2\omega_0^2
\frac{l_1}{l}.
\]
In practice it is convenient to use
\[
\omega_1(eV)=0.40344 \omega_0^2(eV)l_1(\mu m),
\]
where the values $\omega_1, \omega_0$ are expressed in $eV$
and the value $l_1$ is in $\mu m$.

Interrelation between values of $\omega_1$ and $\overline{\omega_p}
~(\overline{\omega_p}=(\overline{\omega_p^2})^{1/2})$ is very essential
for the basic features of the TR. Consider first the case
$\omega_1 \gg \overline{\omega_p}$. For this case the equation
(\ref{5}) has solutions for large $n > 2\omega_1/\overline{\omega_p}$ only.
In this situation the function
\begin{equation}
\sin^2\frac{\varphi_1}{2}=\sin^2\frac{\varphi_2}{2}=
\sin^2\left[n\pi \frac{l_2}{l}\left(1-\frac{\omega_n}{\omega} \right) \right]
\label{7}
\end{equation}
oscillates very fast and one can substitute it by the mean value equal to
1/2. In the integral over $y$ in (\ref{1}) represented as the sum of harmonic
for large $n$ one can substitute summation over $n$ by integration
\begin{equation}
\int_{0}^{\infty}d(\Delta n)~\frac{\sin^2 \left(N\pi (\Delta n)/2\right)}
{\sin^2 \left(\pi (\Delta n)/2\right)} \simeq \frac{2}{\pi} \int_{0}^{\infty}
\frac{dx}{x^2} \sin^2 (Nx)=N
\label{8}
\end{equation}
After this operation the variables $y$ and $\omega$ in (\ref{1})
become independent. This means together with (\ref{8}) that in this case
the TR is the noncoherent sum of the single-interface
contributions (the total
number of interfaces is $2N$). Actually just this case
($\omega_1 \gg \overline{\omega_p}$) is used in the TR
detectors, where the radiated energy is
\begin{equation}
E=\frac{2N}{3}\alpha \omega_0 \gamma
\label{8a}
\end{equation}

In the present paper we consider the opposite case
$\omega_1 \ll \overline{\omega_p}$. In this case the characteristic angles
of radiation are large comparing with $1/\gamma$ (except boundaries of
spectra for the given harmonic). These angles are defined by Eq.(\ref{6})
\begin{equation}
y_n \simeq \frac{\overline{\omega_p^2}}{\omega_n^2}
\left(1-\frac{\omega_n}{\omega} \right) \frac{\omega_n}{\omega},\quad
y_n^{max}=\frac{1}{4}n^2 a,\quad
a=\frac{\overline{\omega_p^2}}{\omega_1^2} \gg 1.
\label{9}
\end{equation}

If one performs collimation of the emitted radiation
with the collimation angle
$y_c=\vartheta_c^2 \gamma^2 < y_n^{max}$ the frequency interval for the
n-th harmonic is split into two parts:
$\omega_n \leq \omega \leq \omega_n^{(1)},~ \omega \geq \omega_n^{(2)}$,
where $\omega_n^{(1,2)}$ are defined by equations following from (\ref{9})
\begin{equation}
\frac{\omega_n}{\omega_n^{(1,2)}}=\frac{1}{2}\left(1 \pm \sqrt{1-z_n} \right),
\quad z_n=\frac{y_c}{y_n^{max}}=\frac{4 y_c}{n^2 a}=
\frac{l}{l_1}\left(\frac{2\vartheta_c \omega_1}{n \omega_0} \right)^2,
\label{10}
\end{equation}
where $\omega_1$ is defined in (\ref{6}). When the collimation is strong
($y_c \ll y_1^{max}= a/4$) the radiation is concentrated in a rather
narrow frequency band
\begin{eqnarray}
&& \Delta \omega_n^{(1)} = \omega_n^{(1)}-\omega_n \simeq \omega_n
\frac{z_n}{4}=\omega_n\frac{y_c}{n^2 a} = \omega_1 \frac{y_c}{n^3 a},
\nonumber \\
&& \omega_n^{(2)}=\frac{4\omega_n}{z_n}=\omega_n \frac{n^2 a}{y_c}
=\omega_1 \frac{n a}{y_c},\quad \frac{y_c}{a}=\frac{l}{l_1}
\frac{\vartheta_c^2 \omega_1^2}{\omega_0^2}.
\label{11}
\end{eqnarray}
The spectral distribution of the radiated energy on the $n$-th harmonic
we obtain integrating Eq.(\ref{1}) over $y$ at $N \gg 1$ in the interval
near $y$ satisfying the equation $\varphi(y)=2\pi n$
\begin{equation}
\int_{}^{}\frac{\sin^2 \left(N \Delta \varphi/2\right)}
{\sin^2 \left(\Delta \varphi/2\right)} dy \simeq
\frac{4\gamma^2}{\omega l}\int_{-\infty}^{\infty} \frac{\sin^2(Nx)}{x^2}dx
=\frac{4\gamma^2}{\omega l}N\pi.
\label{12}
\end{equation}
As it was indicated above, the relative width of the integration interval
$\Delta y/y \sim 1/N$ and within this accuracy at $N \gg 1$ one can use the
formula
\begin{equation}
\frac{\sin^2 \left(N \varphi/2\right)}
{\sin^2 \left( \varphi/2\right)} \simeq  \sum_{n=1}^{\infty}
2 \pi N \delta(\varphi-2\pi n).
\label{13}
\end{equation}
This formula is exact at $N \rightarrow \infty$.

There is some analogy between radiation considered in this paper and the
undulator radiation (see, e. g. \cite{7a}). In undulator the deviation of
the particle's velocity ${\bf v}$ from its mean value varies periodically
under influence of the periodical (in space) magnetic field. In the considered
case such deviation occurs with the wave vector ${\bf k}$
of emitted radiation
(the refraction index) while the particle's velocity remains constant.
However, since the velocity ${\bf v}$ and the wave vector ${\bf k}$
are contained in expressions describing the radiation in the combination
\begin{equation}
1-{\bf n v} \simeq \frac{1}{2} \left(\frac{1}{\gamma^2}+\vartheta^2
+ v_{\perp}^2 +\frac{\omega_0^2}{\omega^2}\right),\quad {\bf n}=
\frac{{\bf k}}{\omega},
\label{13a}
\end{equation}
the both effects are formally equivalent with respect to the coherence of
radiation from different periods. The essential difference
between considered the TR and the undulator radiation is due to the
dependence of Eq.(\ref{13a}) on $\omega$. This property leads to
the following consequences:
\begin{enumerate}
\item the characteristic frequencies of the undulator radiation are directly
proportional to $n$ (in contrast to $\omega_n$ Eq.(\ref{6}))
and inversely proportional to the structure period;
\item in the case considered for fixed $n$ there is the maximal angle
of radiation at $\omega=2\omega_n$ and for smaller angles
$\vartheta < \vartheta_m$ the interval of the allowed frequencies
($\omega \geq \omega_n$) is divided into two intervals, which
are larger than $\omega_n$ in contrast to the undulator case.
\end{enumerate}

Substituting Eqs.(\ref{13})
and (\ref{6}) into (\ref{3}) we obtain for the spectral distribution of
probability of radiation
\begin{eqnarray}
&& \hspace{-10mm}\frac{dw}{d\omega}(y \leq y_c)
=\sum_{n=1}^{\infty} \frac{dw_n^c}{d\omega},\quad
\frac{dw_n^c}{d\omega}=\frac{4\alpha N}{\pi n}
\nonumber \\
&& \hspace{-10mm} \times
\frac{\displaystyle{\sin^2\left[n \pi\frac{l_2}{l}
\left(1-\frac{\omega_n}{\omega}
 \right) \right]}}{(\omega-\omega_n)\displaystyle{\left[1+\frac{l_1}{l}
\left(\frac{\omega}{\omega_n}-1 \right) \right]^2}}
\vartheta(\omega-\omega_n)\left[\vartheta(\omega_n^{(1)}-\omega)+
\vartheta(\omega-\omega_n^{(2)}) \right],
\label{14}
\end{eqnarray}
where $\vartheta(x)$ is the Heaviside step function. Here the threshold
values $\omega_n^{(1,2)}$ are defined in Eq.(\ref{10}), the value
$y_c$ is defined by the collimation angle ($\vartheta_c=\sqrt{y_c}/\gamma$).
Note that Eq.(\ref{14}) is independent of the energy of particle.
However, the condition of applicability this equation depends
essentially on the Lorentz factor $\gamma$:
\begin{equation}
\omega_1=\frac{\omega_0^2 l_1}{4\pi} \ll \overline{\omega_p}=
\gamma \sqrt{\frac{l_1}{l}}\omega_0,\quad \gamma \gg
\frac{\omega_0\sqrt{l_1l}}{4\pi}
\label{15}
\end{equation}
The formula (\ref{14}) was obtained in absence of X-ray absorption which
is especially essential in the low frequency region. Taking into account
absorption we find limitation on the number of foils $N \sim N_{ef}(\omega)$
and
\begin{equation}
N_{ef}(\omega)=\frac{1}{\sigma(\omega)l_1}
\left(1-\exp (-N\sigma(\omega)l_1) \right),
\label{16}
\end{equation}
where $1/\sigma(\omega)$ is the X-ray attenuation length at the
frequency $\omega$. We assume that the absorption in a single foil
is small.

For the strong collimation ($y_c \ll a/4$) the spectral
probability of radiation on the $n$-th harmonic can be written as
\begin{eqnarray}
&& \frac{dw_{n1}^c}{d\omega} \simeq \frac{4\alpha N\pi n^3l_2^2}
{\omega_1^2l^2}\left(\omega-\frac{\omega_1}{n} \right),\quad
\omega-\frac{\omega_1}{n} \geq 0,\quad \omega-\frac{\omega_1}{n} \leq
\frac{\omega_1y_c}{n^3a};
\nonumber \\
&& \frac{dw_{n2}^c}{d\omega} \simeq \frac{4\alpha N \omega_1^2l^2}
{\pi n^3 \omega^3l_1^2}\sin^2\left(\pi n \frac{l_1}{l} \right),\quad
\omega \geq n\omega_1\frac{a}{y_c}.
\label{17}
\end{eqnarray}
These probabilities attain the maximal values on the boundaries of the
regions. These values are
\begin{equation}
\frac{dw_{n1}^c}{d\omega}\left(\omega=\omega_n^{(1)} \right) =
\frac{4\alpha N\pi l_2^2 y_c}{\omega_1l^2 a}, \quad
\frac{dw_{n2}^c}{d\omega}\left(\omega=\omega_n^{(2)} \right) =
\frac{4\alpha N l^2}{\pi \omega_1 n^6l_1^2}\left(\frac{y_c}{a} \right)^3
\sin^2 \left(\pi n\frac{l_1}{l} \right).
\label{18}
\end{equation}
Integrating (\ref{17}) over $\omega$ we obtain the total number
of the passing through the collimator photons:
\begin{equation}
w_{n1}^c =
2\alpha N\pi\frac{1}{n^3} \left(\frac{l_2}{l} \right)^2
\left(\frac{y_c}{a} \right)^2 , \quad
w_{n2}^c =
\frac{2\alpha N}{\pi n^5} \left(\frac{l}{l_1} \right)^2
\left(\frac{y_c}{a} \right)^2
\sin^2 \left(\pi n\frac{l_1}{l} \right).
\label{19}
\end{equation}
The corresponding expressions for the energy losses due to collimated X-ray
emission are
\begin{equation}
E_{n1}^c =
2\alpha N\pi\frac{\omega_1}{n^4} \left(\frac{l_2}{l} \right)^2
\left(\frac{\Delta \omega}{\omega_1} \right)^2 , \quad
E_{n2}^c =
\frac{4\alpha N \omega_1}{\pi n^4}\frac{y_c}{a}
\left(\frac{l}{l_1} \right)^2
\sin^2 \left(\pi n\frac{l_1}{l} \right).
\label{19a}
\end{equation}
It is seen from these expressions that one can neglect the contribution
of higher harmonics ($n \geq 2$). The contribution of the first harmonic
is concentrated in the narrow frequency interval $\Delta \omega \sim
\omega_1 y_c/a \ll \omega_1$ while in the second available interval
the frequencies much larger than $\omega_1$ contribute ($\omega \sim
\omega_1 a/y_c \gg \omega_1$). So, the collimated TR in the forward direction
have quite good monochromaticity.

Note that side by side with absorption there is another process
which imposes limitation on the number of foils $N$ especially when
the angle of collimation is small. This process is multiple scattering
of the projectile which leads to a "smearing" of the $\delta$-function
in Eq.(\ref{13}) and thus reduces the degree of the monochromaticity.
Below we discuss these both processes for some particular examples.

We estimate now the contribution of higher harmonics ($n \gg 1$)
into the spectral distribution of probability in absence of the collimation.
For $\omega \geq \omega_1$ this contribution is suppressed as $1/n^3$.
For $\omega \ll \omega_1$ the main contribution give values
$n \sim \omega_1/\omega \gg 1$. In this situation the spectrum becomes
quasicontinuous, the factor $\sin^2 (\varphi_1/2)$ in Eq.(\ref{1}) is
oscillating very fast and one can substitute it by 1/2. The summation over
$n$ can be replaced by the integration and this is integration of
$\delta$-function in (\ref{13}). For integration over $y$ the
variables $\omega$ and $y$ can be considered as independent.
Taking this into account and integrating (\ref{1}) over interval
$\displaystyle{\frac{4\gamma^2}{\omega l_2} \leq y}~(\frac{\varphi_2}{2}
\geq 1)$ we find
\begin{equation}
\frac{dw}{d\omega}\left(\omega \ll \omega_1 \right) \simeq
\frac{2\alpha N}{\pi \omega}\left[\ln \left(\pi
\frac{\omega_1 l_2}{\omega l_1} \right)+const \right]
\label{20}
\end{equation}

\section{Discussion}

Use of the TR as a source of X-rays was investigated recently in many papers
(see \cite{7}-\cite{12} and references cited therein). Starting from
differential spectral-angular distribution of radiated energy Eq.
(\ref{1}) authors analyzed position of maxima in this distribution.
The corresponding resonance conditions are
\begin{equation}
\frac{\varphi_1}{2}=(2m-1)\frac{\pi}{2},\quad
\frac{\varphi}{2}=m'\pi,\quad m,m'=1,2,3 \ldots
\label{21}
\end{equation}
When these conditions are fulfilled there is connection between the emission
angle and the photon energy.
Measurements were preformed for variety of energies using different
foils and various $N,l_1,l_2$.
Typically spectral X-ray intensity was measured as function
of the emission angle for fixed photon energy or as function of the
photon energy at fixed emission angle.
Experimental results obtained in \cite{7}-\cite{12} are in quite good
agreement with theoretical calculations.

In \cite{11} yield of X-ray photons (with energy $2 \div 6$ keV)
from electrons with energy $\epsilon=855$~MeV
was $N_{\gamma} \sim 10^{-4}$ per electron
and width of the spectral band
was $\Delta \omega \sim \omega$ for Kapton
foils and $N=3$. In \cite{12} yield of X-ray photons
from electrons with energy $\epsilon=900$~MeV
(with energy $14.4$ and $35.5$~keV)
was $N_{\gamma} \sim 2 \cdot 10^{-5}$ and $N_{\gamma} \sim 6 \cdot 10^{-5}$
per electron respectively and
width of the spectral band (FWHM) was $\Delta \omega = 0.5$~keV and
$\Delta \omega = 0.81$~keV
for Silicon monocrystalline foils and $N=10$ and $N=100$.

Now we turn to some specific features of proposed approach.
The ratio of thicknesses $l_1$ and $l_2$ is the important characteristic
of the TR radiator.
The thickness $l_1$ is defined by the radiation frequency
(energy) $\omega_1$ Eq.(\ref{6}) which can be written in a form
\begin{equation}
\omega_1=\alpha m n_e \lambda_c^2 l_1,
\label{22}
\end{equation}
where $n_e = Z n_a$ ($n_a$ is density of atoms in the foil),
$n_e$ is defined in Eq.(\ref{2}),
$\lambda_c=1/m=(\hbar/mc)$ is the Compton wavelength. It is seen from
Eq.(\ref{19}) that the number of collimated photons increases with
$l_2$ (the factor $(l_2/l)^2$). From the other side, the inequality
$\overline{\omega_p} \gg \omega_1$, which have to be fulfilled in our case,
becomes more strong if $l_1/l$ increases. In this situation the requirement
on the collimation angle $\vartheta_c$ becomes more weak and influence of
the multiple scattering diminishes (for the given monochromaticity of
the radiation):
\begin{equation}
\vartheta_c^2 =\frac{l_1}{l}\frac{\omega_0^2}{\omega_1^2}
\frac{\Delta \omega_1}{\omega_1} > \vartheta_s^2 =
\frac{4\pi N l_1}{\alpha \gamma^2 L_{rad}},
\label{23}
\end{equation}
where $L_{rad}$ is the radiation length. So, the optimal value of $l_2$
should be of the same order as $l_1$. Note that in the TR detectors where
$\overline{\omega_p} \ll \omega_1$ there is no limitation connected with
the multiple scattering and the thickness $l_2$ usually one order
of magnitude larger than $l_1$. From (\ref{23}) we have the limitation
on the value $N$ due to multiple scattering
\begin{equation}
N < N_s=\frac{\alpha}{4\pi}\frac{L_{rad}}{l_1}
\frac{\overline{\omega_p^2}}{\omega_1^2}\frac{\Delta \omega_1}{\omega_1}.
\label{24}
\end{equation}
Using the definitions $\overline{\omega_p^2}$ and $\omega_1^2$ (\ref{6})
and explicit formula for $L_{rad}$ (valid for large $Z$) we rewrite
(\ref{23}) in the form
\begin{equation}
N_s=\frac{\gamma^2 l_1}{4 l}
\frac{n_a}{\omega_1^3 \ln \left(183Z^{-1/3} \right) }
\frac{\Delta \omega_1}{\omega_1}.
\label{25}
\end{equation}
So $N_s \propto \gamma^2$ and $N_s \propto \omega_1^{-3}$ and increases
when the energy $\omega_1$ diminishes. For low $\omega_1$ the photon
absorption (see Eq.(\ref{16})) becomes essential. Since for
$\omega \leq 10$~keV the x-ray attenuation length
$1/\sigma(\omega)$ for the heavy elements is from two to three order of
magnitude shorter than for the light elements, one can use in
this region of $\omega$ the light elements only. In the region
$\omega \geq 30$~keV the attenuation length is rather long,
but for large $\omega_1$ the influence of the multiple scattering of the
projectile (\ref{25}) becomes more essential. In this region one can
provide enough large $N_s$ having used the large Lorentz factor only.
For hard X-ray the difference between use of the heavy elements
or the light elements is not so significant.

We consider a few examples of the X-ray yield
basing on the results obtained above.
In Fig.1 the spectral distribution of the radiated energy
$\displaystyle{\frac{d\varepsilon}{d\omega}}$~Eq.(\ref{1})
is shown for Li foils ($\epsilon=25$~GeV, $l_1=0.13~mm,
l=3l_1, N=100$). The important effect of the dependence of the
width and form of the first harmonic
peak ($\displaystyle{\omega_1=\frac{\omega_0^2l_1}{4\pi}=10}$~Kev)
Eq.(\ref{6}) on the collimation angle $\vartheta_c,
(y_c=\vartheta_c^2\gamma^2)$ is given. It is seen that the
collimation is cutting out the different parts of the radiation spectrum.
The connection between the frequency and the emission angle is
the specific property of the undulator type radiation.
When the collimation angle decreases the spectral peak becomes more
narrow (correspondingly the total number of emitted photons
decreases also) and vise versa. The curve 5 exhibits the situation
when the value $y_c$ is near the boundary value $y_{max}$ Eq.(\ref{9}).
For the curve 6 one has $y_c > y_{max}$ and the spectral curve is independent of
the collimation angle. The position of the right slope of the peak
$\omega_n^{(1)}$ is defined from Eq.(\ref{10}), e.g. for $y_c=250$ one
has $\omega_1^{(1)}=12.6$~keV or $\Delta \omega/\omega \simeq 0.25$.

In Fig.2(a) the dependence of the height the
$\displaystyle{\frac{d\varepsilon}{d\omega}}$~Eq.(\ref{1})
on the number of foils ($N=20, 100, 200$) for the fixed collimation angle
($y_c=250$) is presented. The width of the peak
is practically independent of $N$, while the height is proportional to
$N$.The peaks of higher harmonics are situated at $\omega_n=\omega_1/n$
($n$=2, 3, 4). Their characteristics are: the height $\propto 1/n$ and the
width $\propto 1/n^3$ (see Eqs.(\ref{17}))-(\ref{19}), it is necessary
to remind that in Figs.1,2 the energy losses spectrum is
shown, while these equations are for the probability spectrum).
Note that spectra shown in Fig.2(a) are calculated without absorption.
If one takes the absorption into account the higher harmonics will be
strongly suppressed.

There is also the hard part of the emitted spectrum which is given in
Fig.2(b). This is the contribution of the first harmonic (see Eqs.(\ref{18})
and (\ref{19})). For $y_c=250$ the position of the left slope
$\omega_1^{(2)} \simeq 48$~keV (see Eq.(\ref{10})). As the
collimation becomes more strong the value $\omega_1^{(2)}$
increases. Note, that the absorption is much weaker for the hard part
of the spectrum.

Calculating the area of the peak in Fig.1,2
one can estimate number of radiated photons
per electron $N_{\gamma}$, for $y_c=250$ one has $N_{\gamma} \sim 0.1$  for
N=100. Number of foil $N=500 \div 1000$
is typical in many experiments with the TR detectors, the lithium foils were
used in \cite{13}. Quite good estimation of $N_{\gamma}$
can be obtained using (\ref{19}), if $N < N_s$, substituting $N=N_{ef}$.
Because of limitations connected with the photon
absorption and the multiple scattering of the projectile one have to
use the light elements.

Although results given in Figs.1 and 2 are obtained for the particular
substance and the definite parameters, these results are quite universal
at least qualitatively.

There is an opportunity to install the proposed TR radiator in the electron
storage ring. In this case the number $N$ will be defined by the
storage ring operational regime. Let us note that at $N=$20
the first harmonic peak is quite distinct.

The approach proposed in this paper permits to obtain the yield of
X-ray photons $N_{\gamma} \sim 0.01 \div 1$ per electron and the width
of the spectral band up to $\Delta \omega/\omega \sim 0.1$. Of course,
the yield decreases for a more narrow spectral band.
This means at least two order of magnitude
increase of the yield comparing with
obtained in \cite{12}-\cite{13}.
Some specific examples are given in Table where
$\displaystyle{N_1=\frac{1}{\sigma(\omega_1)l_1}}$ defines the
value $N_{ef}$ for fixed $N$ and $\omega_1$~(\ref{16}). The values $N_s$
are calculated according with Eq.(\ref{24}). If $N_s > N_1$ we choose $N$
slightly larger than $N_1$. In the opposite case we choose $N$
slightly smaller than $N_s$. In the first column the material
of the foil is given, ${\rm CH_2}$ is used for Polyethylene.
The value $N_{\gamma}$ is the number of photons
emitted into collimator for the given values of $\omega_1$ and
$\Delta \omega_1$. All calculations are performed for $l=3l_1$.
The results in Table are in a reasonable agreement with Figs.1,2.

\vspace{0.3cm}
{\bf Acknowledgments}

\vspace{0.2cm}
This work
was supported in part by the Russian Fund of Basic Research under Grant
98-02-17866.

\newpage

{\bf Figure captions}

\vspace{15mm}
\begin{itemize}

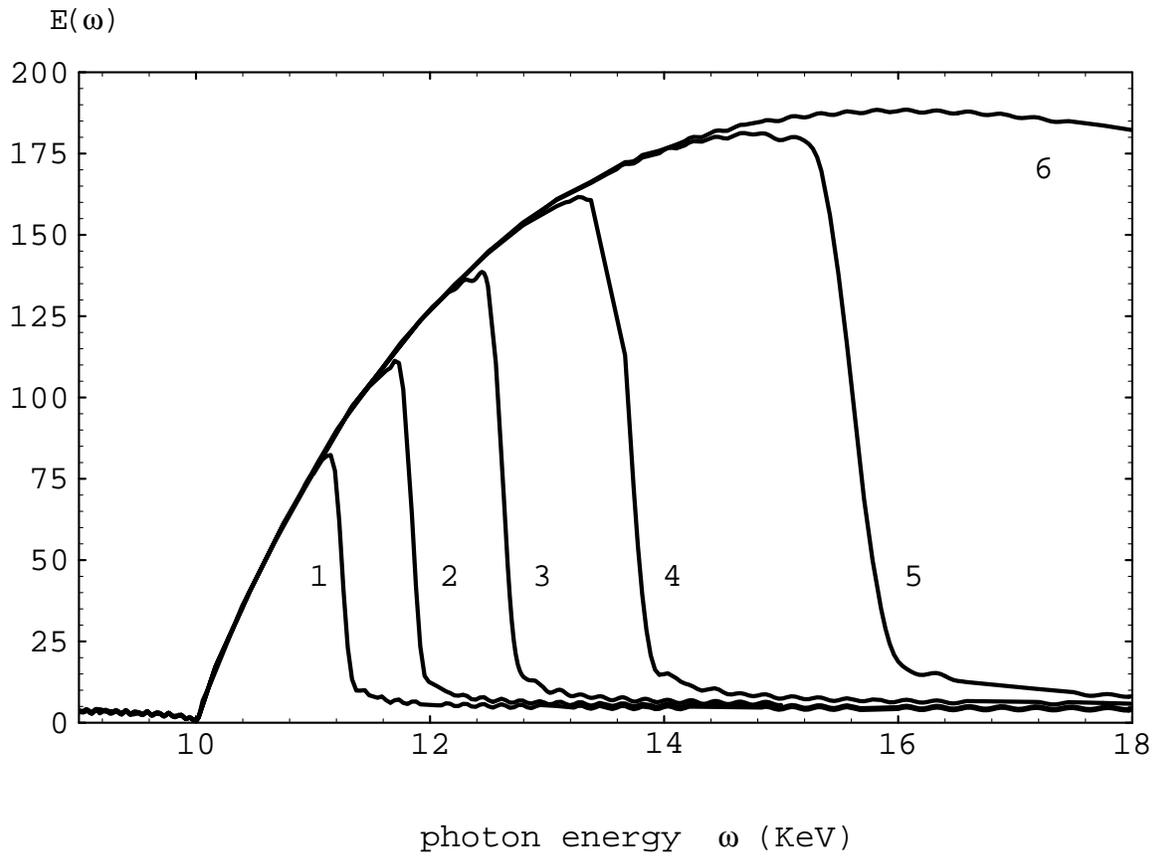
\item {\bf Fig.1} The spectral distribution of the radiated energy
$\displaystyle{E(\omega) \equiv \frac{d\varepsilon}{d\omega}}$~Eq.(\ref{1})
in units $2\alpha/\pi$ vs photon energy
for $Li$ foils ($\epsilon=25$~GeV, $l_1=0.13~mm,
l=3l_1, N=100$). The dependence of the width and form of the first harmonic
peak ($\displaystyle{\omega_1=10}$~Kev) Eq.(\ref{6})
on the collimation angle $\vartheta_c$ is shown.
This angle is measured in units $y_c=\vartheta_c^2\gamma^2$:
$y_c=150, 200, 250, 300, 350, 400$ for curves 1, 2, 3, 4, 5, 6 respectively.

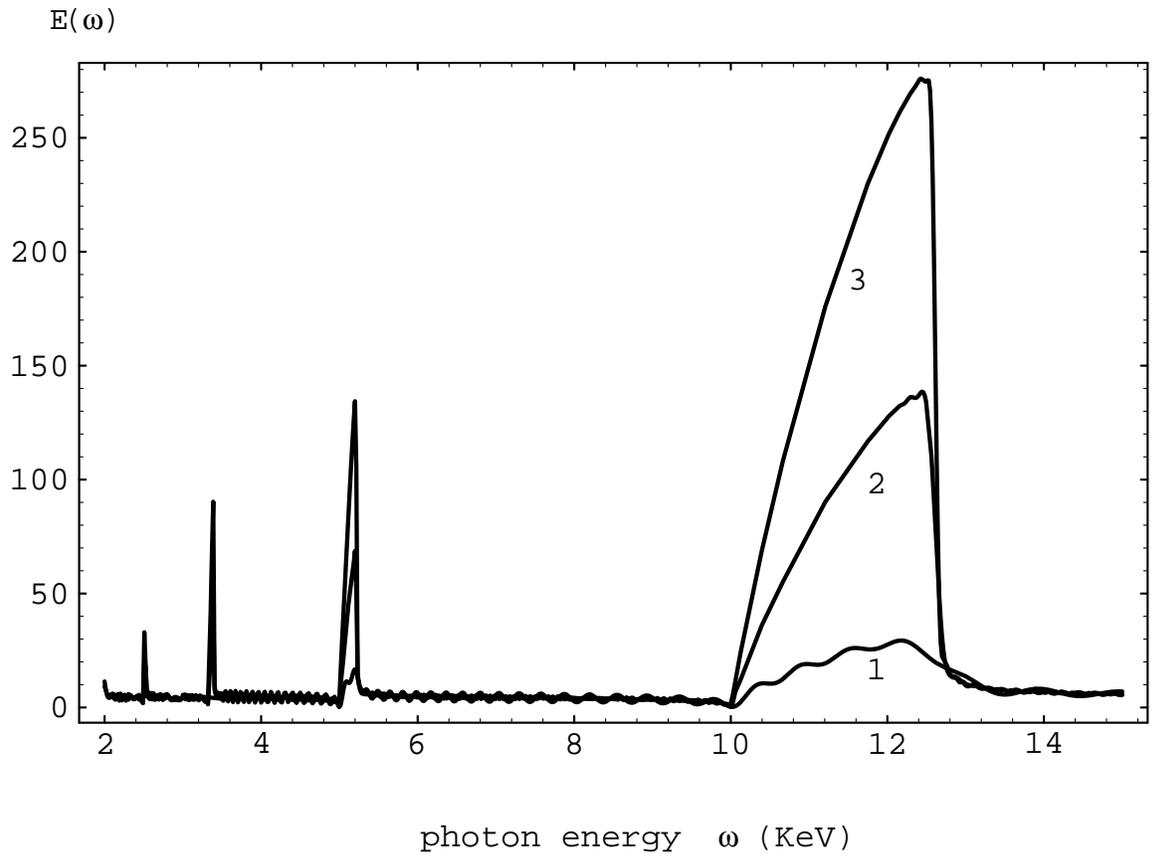
\item {\bf Fig.2} The spectral distribution of the radiated energy
$\displaystyle{E(\omega) \equiv \frac{d\varepsilon}{d\omega}}$~Eq.(\ref{1})
in units $2\alpha/\pi$ vs photon energy
for $Li$ foils ($\epsilon=25$~GeV, $l_1=0.13~mm,
l=3l_1, y_c=250$). The dependence of the height of the spectral curve
on the number of foils is presented ($N=20, 100, 200$
for curves 1, 2, 3 respectively).
\begin{itemize}
\item {\bf (a)} The soft part of the spectral curve.
The main (the first harmonic)
peak and peaks of $n=2, 3, 4$ harmonics ($\omega_n=\omega_1/n$) are seen.

\item {\bf (b)} The hard part of the spectral curve.

\end{itemize}

\end{itemize}

\newpage

\begin{table}
\begin{center}
{\sc TABLE}~
{Characteristic parameters of the radiation process in the TR radiator}\\
\end{center}
\begin{center}
\begin{tabular}{*{11}{|c}|}
\hline
${\rm foil}$ & $\omega_0~(eV)$ & $l_1 (\mu m) $ & $\omega_1~(keV)$ & $N_1$ &
$\varepsilon~(GeV) $& $\displaystyle{\frac{\Delta \omega_1}{\omega_1}}$ &
$N_S~$ & $N$ & $N_{ef}$ & $ N_{\gamma}\cdot 10^2$ \\ \hline
Li & 13.8  & 26 & 2 & 36&  1 & 0.1 & 200& 50  & 27&  0.55 \\ \hline
Li & 13.8  & 39 & 3 & 74&  1 & 0.2 & 120& 100 & 55&  4.5  \\ \hline
Li & 13.8  & 130& 10&440&  5 & 0.3 & 120& 100 & 90&  17   \\ \hline
Li & 13.8  & 130& 10&440&  25& 0.3 &3000& 1000& 400& 72  \\ \hline
Li & 13.8  & 390& 30&360&  25& 0.2 & 74 & 60  & 55 & 4.5  \\ \hline
Li & 13.8  & 650& 50&240&  25& 0.3 & 24 & 20  & 20 & 3.7  \\ \hline
${\rm CH_2}$&20.9  & 57 & 10&110&  5 & 0.2 & 140 & 120& 73 & 6    \\ \hline
${\rm CH_2}$&20.9  & 57 & 10&110&  25& 0.1 & 1750& 200& 92 & 1.9  \\ \hline
${\rm CH_2}$&20.9  & 170& 30&250&  25& 0.2 & 130 & 100& 82 & 6.7  \\ \hline
${\rm CH_2}$&20.9  & 285& 50&190&  25& 0.2 & 28  & 25 & 23 & 1.9  \\ \hline
Si & 31    & 77 & 30&39 &  25& 0.2 & 130 & 80 & 34 & 2.8  \\ \hline
Si & 31    & 130& 30&100&  25& 0.2 & 28  & 25 & 22 & 1.8  \\ \hline
\end{tabular}
\end{center}
\end{table}

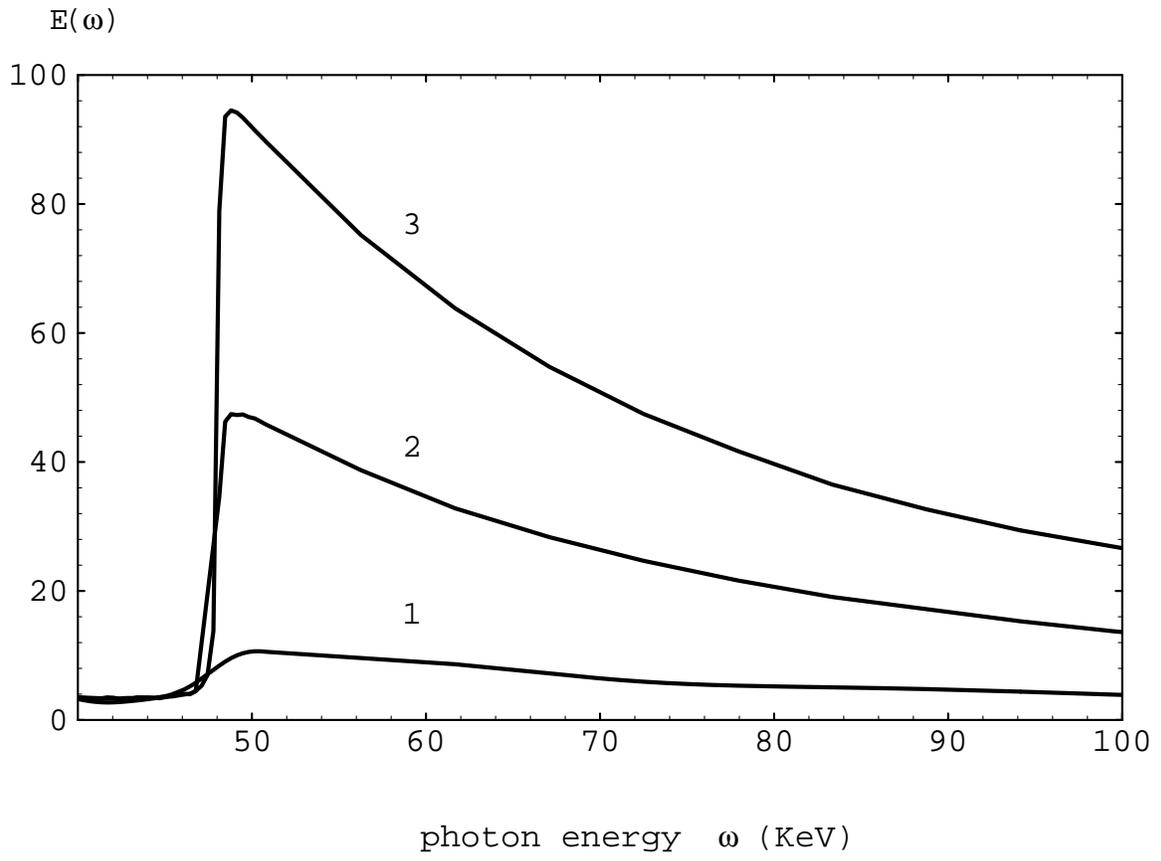
\end{document}